\documentclass[prd,twocolumn,nofootinbib]{revtex4}
\usepackage{graphicx}
\usepackage{latexsym}
\usepackage{amsmath}
\usepackage{amssymb}
\usepackage{epsfig}
\usepackage{fancyheadings}
\usepackage{mathrsfs}
\usepackage{float}

\restylefloat{figure}

\begin{document}

\title{Stability of Cauchy Horizon under perturbations}
\author{Flavio Henrique and Elcio Abdalla}
\affiliation{
Instituto de Fisica, Universidade de Sao Paulo,
  C.P.66.318, CEP 05315-970, Sao Paulo}
\date{February 2004}

\begin{abstract}
We use perturbations in order to study the stability of the Cauchy Horizon
in a Reissner-Nordstr\"om space-time. The perturbations are either scalar
or gravitational, and indicate some strong instabilities.
\end{abstract}

\maketitle
\section{Introduction}

Statistically, most of the Black Holes in the Universe should belong to
the kind described by the Kerr solution, caracterized by the existence of
an external event horizon --- $r_+$ --- and an internal so called Cauchy
horizon, which we denote by --- $r_-$. Classically, an observer can cross
both horizons, and instead of falling into the singularity, emerge into a
new asymptotically flat universe, different from the starting one. The
process can repeat indefinitely, leading the observer to new worlds
\cite{chandra}. 

Such an image is however not a complete description of the physics related
to Black Holes with a Cauchy horizon. Indeed, a small perturbation inside
the horizon can induce an instabilitity which can possibly lead to a
colapse of the geometry.

\begin{figure}[H]
\centering\includegraphics[scale=0.3]{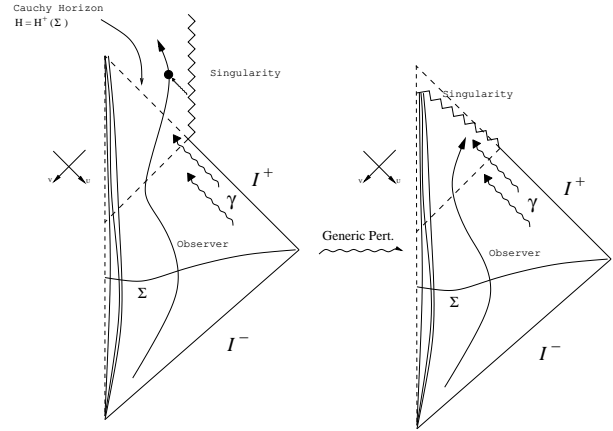}
\caption{{\small Conformal diagram of the Reissner-Nordstr\"om
geometry. Passing through the Cauchy horizon future gets an
indetermination due to the influence of the singularity. Figure from
\cite{penrose}.}}
\label{a-2}
\end{figure}

Moreover, there are further problems connected with the existence of a
Cauchy horizon. The denomination {\it Cauchy horizon} is related to the
fact that depending on how one crosses the surface $r=r_-$ one is outside
the domain of causal dependence on the past.  This means that the Cauchy
horizon is the causal dependence limit $H^{+}(\Sigma)$ of all time-like
curves that intersect a space-like surface $\Sigma$ and cross the Cauchy
horizon, see figure (\ref{a-2}). In other words, the future beyond the
Cauchy horizon cannot be determined exclusively from the past. In
particular, for the Kerr Black Hole the region beyond the internal horizon
can contain curves which are closed in time, see \cite{o'neil}.  These
curves introduce an infinite set of possible histories for any event into
``causal'' influence domain of ``time machines''.

There seem to be difficult points for interpretation beyond the Cauchy
horizon, namely crossing it is {\it fraught with danger} using words of
Chandrasekhar \cite{chandra}. Indeed, the Cauchy horizon is known to be a
structure of great unstability under perturbations whose intensity
enlarges in its neighbourhood. In this work we study the stability of
Cauchy horizon under perturbations \cite{novikov1,novikov2}. Some
non-perturbative results have been obtained by \cite{gnedin} and
\cite{brady2}, where unstability has been found.

In this work we first choose the geometry and study its stability
for a fictitious observer traveling into the horizon in a time-like
curve. Later we develop a scalar and a gravitational perturbation, after
which we arrive at some conclusions.

\section{Geometry and Observer}

It is a difficult task to work with a Kerr geometry, due to the fact that
coordinates do not decouple. An easier alternative in order to study
Cauchy horizons is to consider a Reissner-Nordstr\"om geometry, defined by
the metric
\begin{equation}\label{a}
ds^2=\frac{\Delta}{r^2}dt^2-\frac{r^2}{\Delta}dr^2-r^2d\theta^2-r^2\sin^2
\theta
d\varphi^2\, ,
\end{equation}
where $\Delta=r^2-2Mr+Q^2$.

Besides representing a structure similar to the Kerr, it corresponds to
the limit of the Kerr geometry for low density of angular momentum by
mass, being a reasonable approximation for the problem.

We are interested in the region between the two horizons, where $\Delta
<0$. In this case the variables $r$ and $t$ exchange their role (see
\cite{hawking}). The tortoise coordinates are, in this case, defined as
\begin{align}\label{b}
dt^* &=-\frac{r^2}{\Delta}dr\quad , & dr^* &=dt\quad ,
\end{align}
where the definitions contain the new status of time and distance,
respectively. Notice that $r\rightarrow r_-$ implies $t^{*}\rightarrow
-\infty$.

We now suppose that small perturbations are generated between the two
horizons, with the ``energy density'' measured in the reference frame of the
observer as

\begin{equation}\label{b-1}
\mathscr{F}=p^{\mu}\partial_{\mu}\Psi \quad ,
\end{equation}
as proposed in \cite{novikov1} (see also \cite{chandra}). The $p^\mu$ in
expression (\ref{b-1}) corresponds to the traveller's quadri-momentum in a
free falling time-like curve into Cauchy horizon at the equator of the black 
hole, i. e. $\theta=\frac{\pi}{2}$ and with zero of angular momentum. 
Expression (\ref{b-1}) is a part of traveller's complete energy
density which is 
\[
\mathscr{E}=\mathscr{F}^2-\frac{1}{2}p^2\partial_\mu\Psi\partial^\mu\Psi
\quad , 
\]
where $\Psi$ is the perturbation. In the coordinate frame
\begin{align}\label{c}
u &=t^*-r^*, & v &=t^*+r^*
\end{align}
the ``energy density'' is given by the expression
\begin{multline}\label{d}
\mathscr{F}=-\frac{r^2}{(r_+-r)(r-r_-)}\left[\left(
E+\sqrt{E^2-\frac{\Delta}{r^2}}\right)\Psi,_v \right. \\
\left.-\left( E-\sqrt{E^2 -\frac{\Delta}{r^2}}\right)\Psi,_u\right]\quad ,
\end{multline}
where  $E$ is a real number between $-\infty$ and $+\infty$. Near the
Cauchy horizon $\mathscr{F}$ behaves as
\begin{eqnarray*}
\mathscr{F}_u\rightarrow \frac{r_{-}^2}{r_+-r_-}E
e^{-\frac{\kappa_-}{2}u}\Psi,_u|_{v=v_0}~\text{for}~u\rightarrow
-\infty\, ,\\
\mathscr{F}_v\rightarrow -\frac{r_{-}^2}{r_+-r_-}E e^{-\frac{\kappa_-}{2}v}
\Psi,_v|_{u=u_0}~\text{for}~v\rightarrow -\infty,
\end{eqnarray*}
with $\kappa_- =\frac{r_+-r_-}{r_{-}^{2}}$. Supposing that the asymptotic
behaviour of the perturbation is $\Psi ,_u\rightarrow e^{\alpha
u}$ and $\Psi ,_v\rightarrow e^{\alpha v}$; one concludes that for
$\alpha-\frac{\kappa_-}{2}<0$ the perturbation is unstable since near the
Cauchy horizon $|\mathscr{F}|\rightarrow \infty$. Otherwise, for $\alpha
-\frac{\kappa_-}{2}>0$ the perturbation is stable. The possibility
$\alpha=\frac{\kappa_-}{2}$ implies convergence of $\mathscr{F}$ to an
asymptotic and finite number, whose amplitude depends of global evolution
of $\Psi$, in which case the plot of $\mathscr{F}$ is necessary.

\section{Scalar Field}

We consider the scalar wave equation
\begin{equation}\label{f}
\frac{1}{\sqrt{-g}}\partial_{\mu}\sqrt{-g}\partial^{\mu}
\Phi(r,t,\theta,\varphi)=0,
\end{equation}
where $g= det\lbrack g_{\mu\nu}\rbrack = -r^2\sin^2\theta$, for the
Reissner-Nordstr\"om metric. We assume the usual separation in terms of
spherical harmonics,
\[
\Phi(r,t,\theta,\varphi)=\sum_{lm}\frac{\Psi
(r,t)}{r}(r,t)Y_{lm}(\theta,\varphi)\quad ,
\]
Obtaining the wave equation
\begin{multline*}
\frac{\Delta}{r^2}\frac{\partial}{\partial
  r}\left(\frac{\Delta}{r^2}\frac{\partial}{\partial
  r}\Psi(r,t)\right)-\frac{\partial^2}{\partial
  t^2}\Psi(r,t)= \\ \frac{\Delta}{r^3}\left(\frac{\partial}{\partial
  r}\frac{\Delta}{r^2}+\frac{l(l+1)}{r}\right)\Psi(r,t)\quad ,
\end{multline*}
which for the tortoise coordinates (\ref{b}) leads to
\begin{equation}\label{g}
\frac{\partial^2}{\partial {t^*}^2}\Psi(r^*,
t^*)-\frac{\partial^2}{\partial {r^*}^2}\Psi(r^*,
t^*)=V_{l}(t^*)\Psi(r^*, t^*)\quad ,
\end{equation}
with $V_l$ given by
$V_l(t^*)=\frac{\Delta}{r^4}\left[\frac{2}{r^2}(Mr-Q^2) + l(l+1)\right]$.
Using the $u,v$ coordinates, we find, as usual,
\begin{equation}\label{h}
4\frac{\partial^2}{\partial u \partial
v}\Psi=V_{l}(\frac{u+v}{2})\Psi\quad .
\end{equation}

\begin{figure}[H]
\centering\includegraphics[scale=0.3]{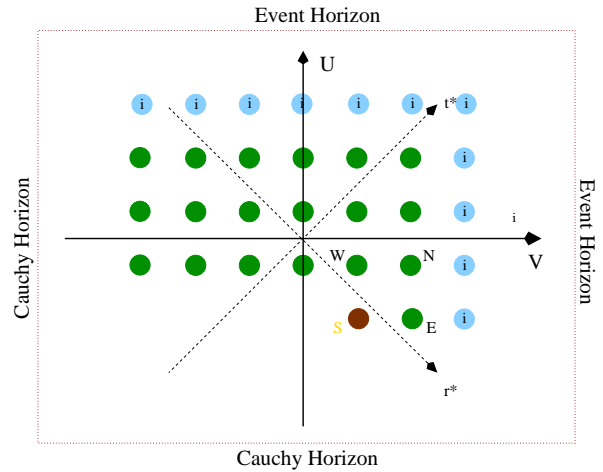}
\caption{{\small Diagram corresponding to the integration grid, in the
region between the two horizons, in the Reissner-Nordstr\"om metric. The
cells marked with $i$  correspond to the inicial condition.}}
\label{i}
\end{figure}

Such equation is discretized as shown in
figure (\ref{i}) and following \cite{brady,abdalla}, whereof we use the
notation.

Therefore,
\[
\Psi_S=(\Psi_W+\Psi_E)\left(1+\frac{\delta v\delta
u}{8}V(t_{W}^*)\right)-\Psi_N\quad .
\]

\begin{figure}[H]
\centering\includegraphics[scale=1.4]{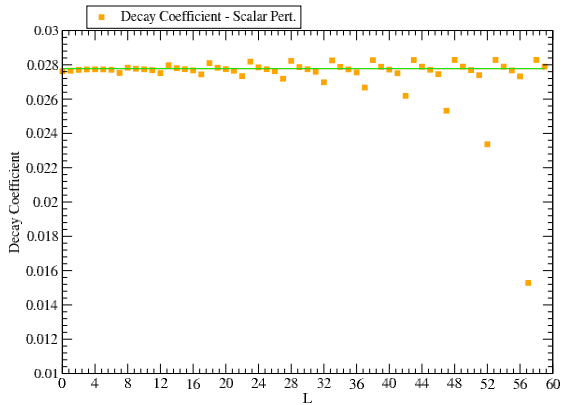}
\caption{{\small $\alpha$ coefficients for $r_+=3.5$ and $r_-=3.0$
from the assimptotic behaviour of  $\Psi ,_u|_{v_0=-300.00}$
for $r\rightarrow r_-$. }}
\label{e1}
\end{figure}
\begin{figure}[H]
\centering\includegraphics[scale=0.45]{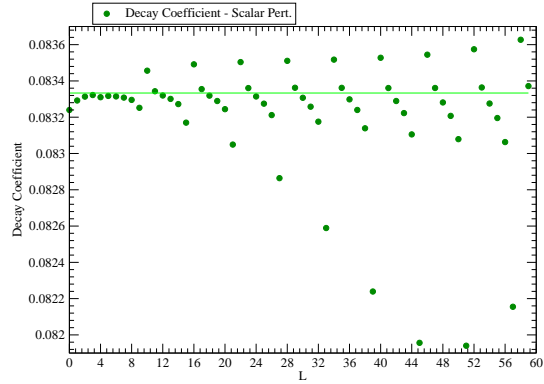}
\caption{{\small $\alpha$ coefficients corresponding to the function
$\Psi ,_v|_{u_0=-140.00}$ for  $r_+=4.5$ and $r_-=3.0$. We used
$\Delta u=\Delta v=180$.}}
\label{e2}
\end{figure}
\begin{figure}[H]
\centering\includegraphics[scale=0.35]{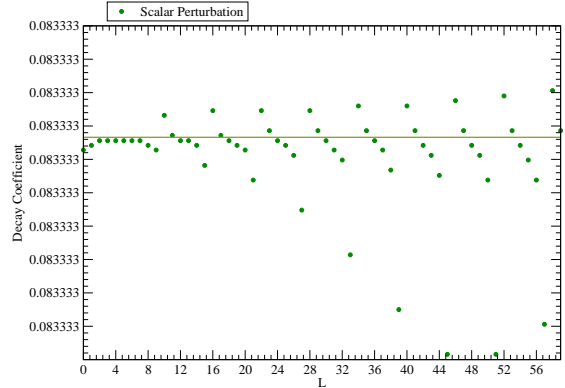}
\caption{{\small $\alpha$ coefficients corresponding to the function
$\Psi,_v|_{u_0=-240.00}$. Here  $\Delta v=\Delta u=280.00$. Notice that
now all $\alpha$'s are very near the asymptotic value
$\frac{\kappa_-}{2}=0,0833$.}}
\label{e3}
\end{figure}

The integration of equation (\ref{h}) has been performed for all harmonics
$l$ from 0 up to 59. The results are shown in figures (\ref{e1}),
(\ref{e2}) and (\ref{e3}). The straight line indicates the transition
between stable perturbations, above the straight line, and unstable below.
For $l$ above 8 the coefficients oscilate between stability and
unstability, a fact observed thereafter. Further observations are here
performed since enlarging the grid towards the horizons leads to the fact
that all perturbations stay on the verge of stability and a more detailed
investigation was required. Indeed, stable points can undergo drastic
variations when one approaches the Cauchy horizon  as shown in figure
(\ref{e3}) which is analogous to figure (\ref{e2}), but with an enlarged grid.

\begin{figure}[H]
\centering\includegraphics[angle=-90,scale=0.3]{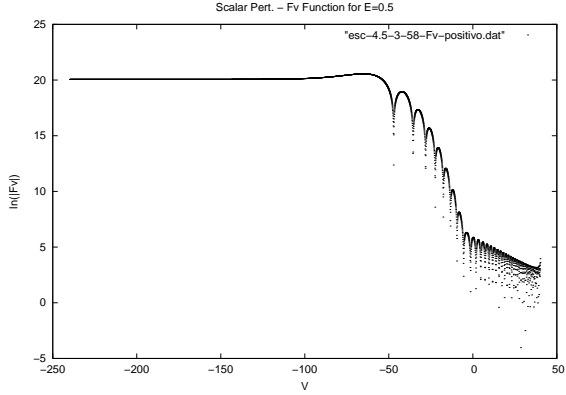}
\caption{{\small Diagram sowing the function $\ln |\mathscr{F}_v|\times v$
for $l=58$. There is no sign of stability.}}
\label{X}
\end{figure}

On example of such a behaviour is shown in figure (\ref{X}) where the
function $\mathscr{F}_v|_{u_0=-240.0}$ was plotted for $l=58$. Although
from figure (\ref{e3}) the $\alpha$ coefficient indicates stability, the
function $\mathscr{F}_v$ reveals a strong increase by 16 orders of
magnitude. Such a behaviour turns out to be quite general.


\begin{figure}[H]
\centering\includegraphics[angle=-90,scale=0.3]{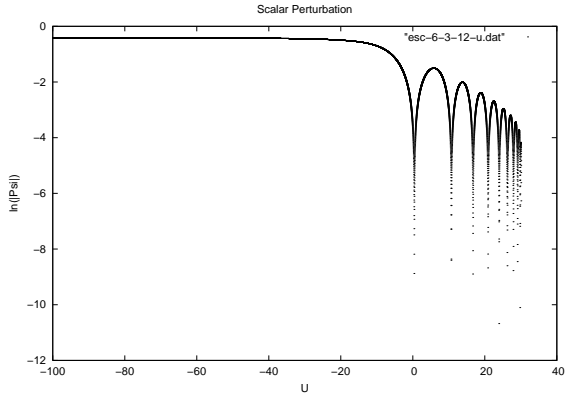}
\caption{{\small Here $\ln |\Psi (u,v=-100)|$ is shown, for a scalar
perturbation for $r_+=6.0$ and $r_-=3.0$ with $l=12$. Between the points
(30.0) and  (-100.0) there is a growth by three orders of magnitude.}}
\label{e4}
\end{figure}

\begin{figure}[H]
\centering\includegraphics[angle=-90,scale=0.3]{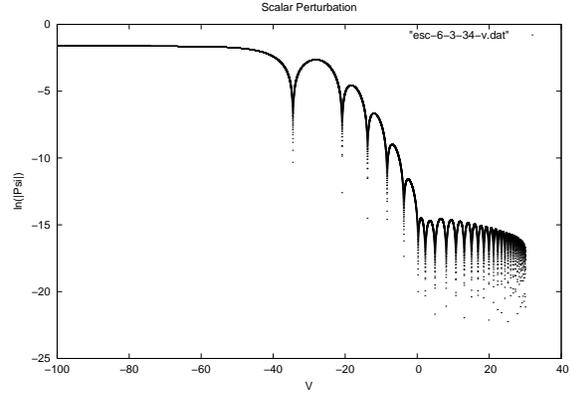}
\caption{{\small Here $\ln |\Psi (u=-100,v)|$   is shown, for a scalar
perturbation for $r_+=6.0$ and $r_-=3.0$ with $l=34$. There is an increase
here in 15 orders of magnitude.}}
\label{e5}
\end{figure}

\begin{figure}[H]
\centering\includegraphics[angle=-90,scale=0.3]{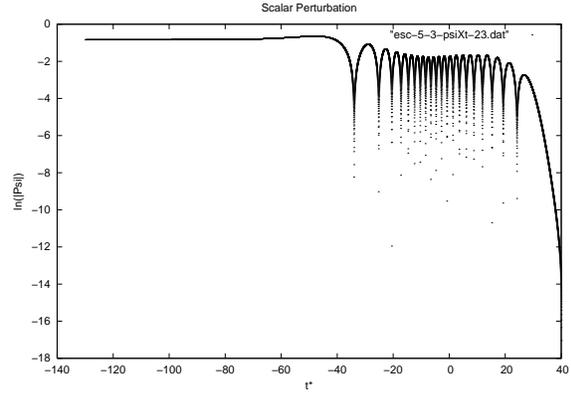}
\caption{{\small Diagram of $\ln |\Psi |\times t^*$ for $r_+=5.0$ and
$r_-=3.0$ with $l=23$. There is a big increase in the amplitude at the
position $r^*=0.0$ as we approach the Cauchy horizon.}}
\label{e6}
\end{figure}

Furthermore, some results concerning the perturbation as function of the
$u$, $v$ and $t^*$ coordinates are shown in figures
(\ref{e4}), (\ref{e5}) and (\ref{e6}), where in those diagrams the Cauchy
horizon corresponds to the far left of the diagrams. These diagrams reveal
important unstabilities torward the Cauchy horizon, with an increase of
three orders of magnitude in figure (\ref{e4}), 15 orders in figure
(\ref{e5}), and 13 orders in figure (\ref{e6}), no matter what is the
choice for $r^*$.

Further examples have been studied with analogous results.

\section{Gravitational Perturbation}

We perturb the metric (\ref{a}), refered to as  $g^{(0)}_{\mu\nu}$, by
means of a small correction $h_{\mu\nu}$ as
\begin{equation}\label{a-1}
g_{\mu\nu}=g_{\mu\nu}^{(0)}+h_{\mu\nu}\quad .
\end{equation}

We shall complement our study now about perturbations on
Reissner-Nordstr\"om geometry using directly the above procedure. The
perturbation (\ref{a-1}) leds to two types of perturbations, as usual
\cite{chandra}, which corresponds to the scalar equation with a different
potential $V_{l}(t^*) $, namely,
\begin{itemize}
\item{axial:}
\end{itemize}
\[
V_{l}^-(t^*)=V_{l}^-(r(t^*))=\frac{\Delta}{r^5}\left[
(\mu^2+2)r-q+\frac{4Q^2}{r}\right]\quad ,
\]
\begin{itemize}
\item{polar:}
\end{itemize}
\[
V_{l}^+(t^*)=V_{l}^+(r(t^*))=V_{l}^- -2q\frac{d}{dt^*}
\frac{\Delta}{r^3(\mu^2r+q)}\quad ,
\]
where $\mu^2=(l-1)(l+2)$ and $q=3M-\sqrt{9M^2+4Q^2\mu^2}$.

Equation (\ref{g}) can again be integrated.

\begin{figure}[H]
\centering\includegraphics[scale=0.45]{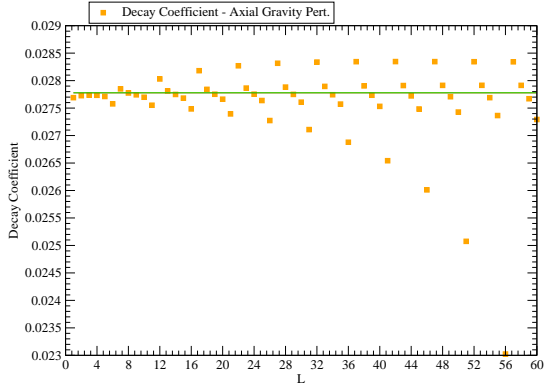}
\caption{{\small  $\alpha$ coefficients for $\Psi
,_u|_{v_0=-300.00}$ in the asymptotic limit $r\rightarrow r_-$ for
$r_+=3.5$ and $r_-=3.0$.}}
\label{g1}
\end{figure}
\begin{figure}[H]
\centering\includegraphics[scale=0.44]{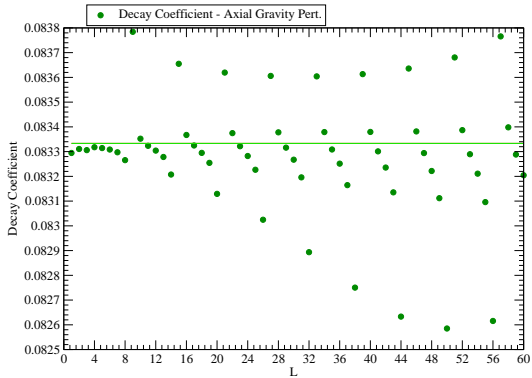}
\caption{{\small $\alpha$'s corresponding to $\Psi
,_v|_{u_0=-140.00}$. Here we used
$\Delta u=\Delta v=180$.}}
\label{g2}
\end{figure}

Figures (\ref{g1}) and (\ref{g2}) give us the results for the  $\alpha 's$
with $l$ from 1 up to 60.

\begin{figure}[H]
\centering\includegraphics[scale=0.45]{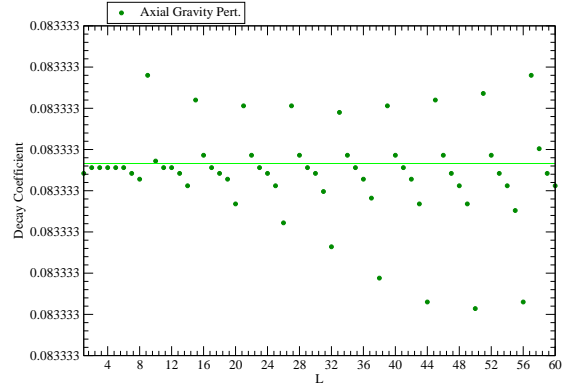}
\caption{{\small $\alpha$'s  corresponding to $\Psi ,_v|_{u_0=-240.00}$
using $\Delta
u=\Delta v=280.00$.}}
\label{g3}
\end{figure}

For $l>8$ there are again oscilations between stable and unstable points.
As in the scalar field we amplify the grid of integration.
Figure (\ref{g3}) shown results similar to those obtained for the scalar
field.

\vspace{0.2cm}

\begin{figure}[H]
\centering\includegraphics[angle=-90,scale=0.3]{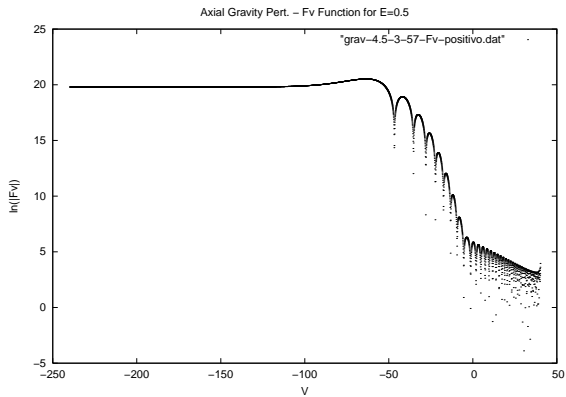}
\caption{{\small $\ln |\mathscr{F}_v|\times v$  for $l=57$. The final
 value of the function is 17 orders of magnitude larger than the inicial
 value.}}
\label{Y}
\end{figure}

Figure (\ref{Y}) displays the function $\mathscr{F}_v|_{u_0=-240.00}$ for
the value $l=57$ from figure (\ref{g3}).

We learn again that in spite of the fact that a rough calculation of
$\alpha$ indicated stability, in fact  $\mathscr{F}_v$ eventually
undergoes an increase of 17 orders of magnitude. The same occurs with any
other point in figure (\ref{g3}). This occurs because the
$\alpha$-coefficient does not stay fixed, converging to
$\frac{\kappa_-}{2}$ when $v\rightarrow -\infty$.

\begin{figure}[H]
\centering\includegraphics[scale=0.37]{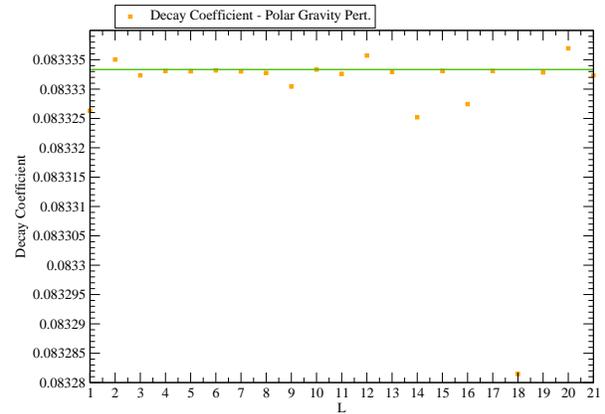}
\caption{{\small  $\alpha$ coefficients for the polar perturbation
$\Psi ,_u|_{v_0=-190.00}$ when $r_+=4.5$ and
$r_-=3.0$.}}
\label{g4}
\end{figure}

\begin{figure}[H]
\centering\includegraphics[scale=0.37]{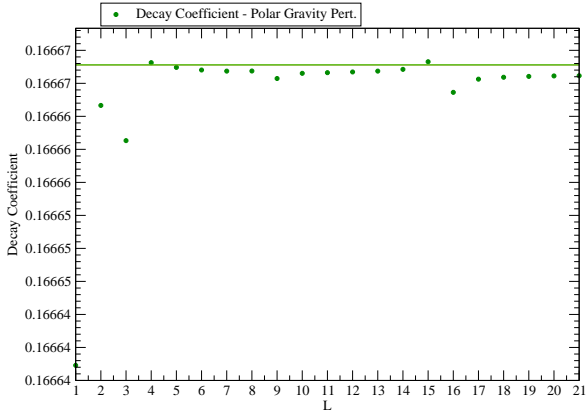}
\caption{{\small $\alpha$ coefficients for the polar perturbation $\Psi
,_v|_{u_0=-120.00}$ with  $r_+=6.0$ and $r_-=3.0$.}}
\label{g5}
\end{figure}

Figures (\ref{g4}) and (\ref{g5}) shown the $\alpha$ coefficients for the
polar perturbation for $l=1,\ldots ,21$. In contrast to the axial perturbation
the coefficients vary little and with no periodicity,

\begin{figure}[H]
\centering\includegraphics[angle=-90,scale=0.3]{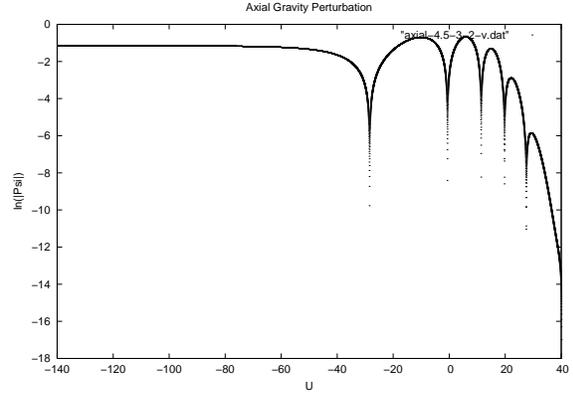}
\caption{{\small $|\Psi (u=-140,v)|\times v$ for $r_+=4.5$ and
$r_-=3.0$ with $l=2$.}}
\label{g6}
\end{figure}

\begin{figure}[H]
\centering\includegraphics[angle=-90,scale=0.3]{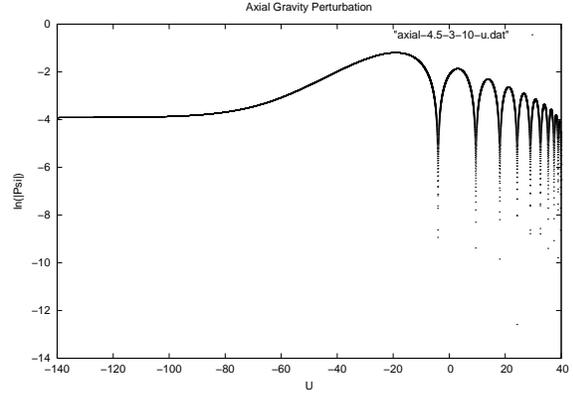}
\caption{{\small $|\Psi (u,v=-140)|\times u$ for
$r_+=4.5$ and $r_-=3.0$ with $l=10$.}}
\label{g7}
\end{figure}

\begin{figure}[H]
\centering\includegraphics[angle=-90,scale=0.3]{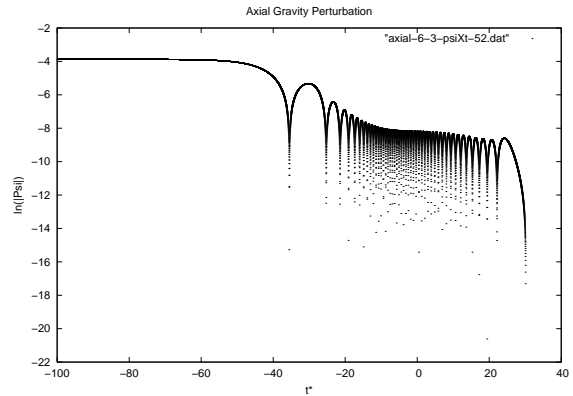}
\caption{{\small $\ln |\Psi | \times t^*$ for $r_+=6.0$
and $r_-=3.0$ with $l=52$. Instability of the perturbation at $r^*=0.0$.}}
\label{g8}
\end{figure}

\begin{figure}[H]
\centering\includegraphics[angle=-90,scale=0.3]{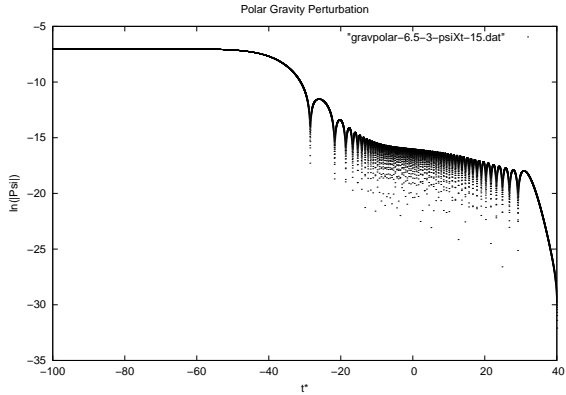}
\caption{{\small $\ln |\Psi |\times t^*$  $r_+=6.5$ and $r_-=3.0$ with
 $l=15$.}}
\label{g11}
\end{figure}

\begin{figure}[H]
\centering\includegraphics[angle=-90,scale=0.3]{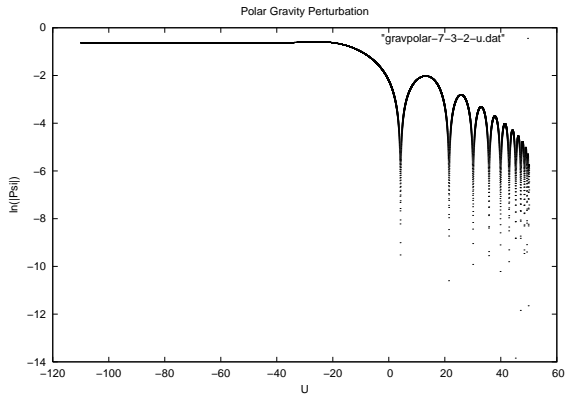}
\caption{{\small Polar perturbation $\ln |\Psi (u,v=-115)|$ for $r_+=7.0$
and $r_-=3.0$ with $l=2$.}}
\label{g9}
\end{figure}

\begin{figure}[H]
\centering\includegraphics[angle=-90,scale=0.3]{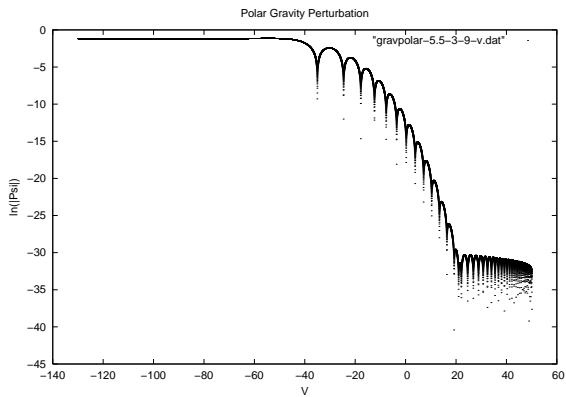}
\caption{{\small Polar perturbation  $\ln |\Psi
(u=-130,v)|$ for $r_+=5.5$ and $r_-=3.0$ with $l=9$.}}
\label{g10}
\end{figure}

We also show the perturbations of the functions $V_{l}^{\pm}$ in figures
(\ref{g6}), (\ref{g7}), (\ref{g8}), (\ref{g11}), (\ref{g9}) and
(\ref{g10}). The results are similar to those obtained for scalar
perturbations, with large increase of the functions tword the Cauchy
horizon. Notice that polar perturbation has the most unstable results when
compared with other equivalent diagrams.

As a conclusion, polar perturbation shows even more enhanced instability.

\section{Conclusion}

The results obtained for the parameters $\alpha$ reveal some regions of
stability. However, as we approach the Cauchy horizon these coefficients
vary, eventually converging to  $\frac{\kappa_-}{2}$. It is thus necessary
to analyse the density of energy $\mathscr{F}$, as shown in figures
(\ref{X}) e (\ref{Y}), which in spite of corresponding to $\alpha$
parameters in the region of stability.

We also show diagrams for $\ln |\Psi|\times u$, $\ln |\Psi|\times v$ and
$\ln |\Psi|\times t^*$, showing large instabilities. 

The final picture is that towards the Cauchy horizon a large instability
occurs, and the energy density tends to increase with no bound. It is thus
possible the space-time itself gets corrected in a nonperturbative
way. Thus a passage by the Cauchy horizon may be impossible, a way o
preventing several possible difficulties with different Universes or
problems with causality, in an example of Hawking's  conjecture of
cronological protection \cite{hawking2}. Perturbation theory of the old
style fails in contrast to previous results, \cite{ori-2}, \cite{gnedin}
\cite{brady2} and new non perurbative methods are mandatory
\cite{wangabdlin}.

\section{Acknowledgements}

We are grateful to Dr. Carlos Molina for the assistance in the use of
numerical codes. This work has been supported by FAPESP and CNPq
(Brazil).

\end{document}